# Velocity anisotropy and the antiferromagnetic instability in the $QED_3$ theory of underdoped cuprates

Dominic J. Lee and Igor F. Herbut

*Department of Physics, Simon Fraser University, Burnaby, British Columbia, Canada V5A 1S6*

(Dated: November 9, 2018)

We consider the effective 2+1 dimensional electrodynamics ($QED_3$) of low-energy quasiparticles coupled to fluctuating vortex loops in the d-wave superconductor, with the velocity (Dirac) anisotropy $v_F/v_\Delta \neq 1$. This theory should be relevant to the quantum superconductor-insulator transition in underdoped cuprates. Working in the customary large-N approximation, we find that weak anisotropy is a marginally irrelevant perturbation to the Lorentz invariant $QED_3$, and that the critical number $N_c$ of Dirac fields below which the theory suffers the antiferromagnetic (chiral) instability stays the same.

Two states of matter ubiquitous in all high-temperature superconductors are the insulating antiferromagnet and the d-wave superconductor. Recently one of us has shown [1] that, while experience has taught us that to derive superconductivity from antiferromagnetism is rather difficult, the opposite appears possible: d-wave superconductors (dSC) at $T = 0$ have an inherent instability towards antiferromagnetism once they loose their phase coherence. The low-energy theory that describes this quantum phase transition is the anisotropic 2+1 dimensional quantum electrodynamics ($QED_3$), in which the *two* ($N = 2$) Dirac fermions correspond to neutral spinons near the two pairs of diagonally opposed nodes in the d-wave state, and the massless gauge field [2] describes the interaction with the unbound vortex loops. Such a theory in the isotropic limit is well known to have a "chiral" instability for the number of Dirac components $N < N_c$, with $N_c \approx 3$ [3]. In the context of d-wave superconductivity, this instability corresponds to the dynamically generated staggered potential felt by the original electrons, i.e. a weak spin density-wave (SDW) order [1]. The SDW obtained this way is then enhanced by Coulomb repulsion, and is assumed to be smoothly connected to the commensurate antiferromagnet right at half filling. Another crucial result of this approach is that such spontaneously generated SDW automatically confines neutral spin-1/2 excitations (spinons) of the superconductor. If, on the other hand, the critical number of components were $N_c < 2$, the ground state would be qualitatively different: the phase incoherent dSC would not show any kind of order at $T = 0$, and the ground state would be analogous to the "algebraic Fermi (spin) liquids" [2], [4], invoked in literature to describe the pseudogap regime in cuprates. Evidently, it is of some importance to determine in which way the perturbations to the Lorentz invariant $QED_3$, relevant to cuprates, affect the critical number of components for the SDW instability.

One such perturbation that is quite significant in high-temperature superconductors is the large observed difference between the two characteristic velocities that appear in the linearized Hamiltonian for the d-wave quasiparticles, $v_F/v_\Delta \sim 10$ [5]. Here $v_F$ is the Fermi velocity, and $v_\Delta$ is related to the amplitude of the superconducting order parameter. Similar velocity anisotropy also appears in the other incarnations of the $QED_3$ in the theories of high-$T_c$ superconductivity [6], and is in fact generic to the effective Dirac theories describing condensed matter systems [7], which have no reason to be Lorentz invariant. Because of the coupling to the gauge-field such anisotropy can not be simply rescaled out, and is in fact precisely a marginal perturbation to the Lorentz invariant $QED_3$ by simple power counting. In this paper we study the effect of weak anisotropy of this type on the critical number of components $N_c$ in the effective $QED_3$.

Working in a large-N approximation, we find two results. Firstly, that weak anisotropy becomes *irrelevant* to the leading order in $1/N$. Secondly, we argue that $N_c$ is the same in the weakly anisotropic case as it is in the isotropic case.

We calculate both the renormalized anisotropy and $\Delta N_c$ by performing expansions in $\lambda' = \lambda - 1$, $\delta' = \delta - 1$ and $1/N$, where $\lambda = v_F/v_\Delta$ and $\delta = \sqrt{v_F v_\Delta}/c$ (with $c$ being the propagation velocity of the gauge field), around the self-consistent approximation scheme of [3]. Essentially, when we come to calculate $N_c$, this becomes an expansion in powers of $1/N_c^0$ around $N_c^0$, the value of $N_c$ obtained by [3] in the isotropic limit, as well as in powers of $\lambda'$ and $\delta'$. In such an expansion we find that to leading order in $1/N_c^0$ $N_c$ is an *increasing* function of weak anisotropy. Since we find that weak anisotropy is irrelevant to $O(1/N)$, when we rewrite our result in terms of the renormalized anisotropies $\lambda_R$ and $\delta_R$ (to leading order in $1/N_c^0$, this corresponds to setting $\lambda = \lambda_R$ and $\delta = \delta_R$) we find that $N_c = N_c^0$.

The effective low-energy theory for the d-wave quasiparticles (i. e. spinons) at $T = 0$ coupled to the topological defects (vortex loops [8] ) in the superconducting phase can be written as $\mathcal{L} = \mathcal{L}_1 + \mathcal{L}_2 + \mathcal{L}_A$ with [1]:

$$\mathcal{L}_1 = i \sum_j^{N/2} \bar{\Psi}_{j,1} \left[ \gamma_0(\partial_0 - ieA_0) + \frac{\delta}{\sqrt{\lambda}} \gamma_1(\partial_1 - ieA_1) \right.$$
$$\left. + \delta\sqrt{\lambda} \gamma_2(\partial_2 - ieA_2) \right] \Psi_{j,1}$$
$$\mathcal{L}_A = (\nabla \times \mathbf{A})^2 + \frac{(\nabla \cdot \mathbf{A})^2}{\epsilon}, \quad (1)$$

and $\mathcal{L}_2$ is the same as $\mathcal{L}_1$, except that $1/\sqrt{\lambda} \leftrightarrow \sqrt{\lambda}$. We rescaled the coordinates so that $\delta = \sqrt{v_F v_\Delta}/c$ and $\lambda = v_F/v_\Delta$. The $\gamma$-matrices are $\gamma_0 = (i/\sqrt{2})\sigma_1 \otimes I$, $\gamma_1 = (i/\sqrt{2})\sigma_2 \otimes \sigma_3$, and $\gamma_2 = -(i/\sqrt{2})\sigma_2 \otimes \sigma_1$, and satisfy the Clifford algebra $\{\gamma_\mu, \gamma_\nu\} = -\delta_{\mu\nu}$. $\vec{\sigma}$ are the standard Pauli matrices. The gauge field $\mathbf{A}$ represents the singular superconducting phase fluctuations induced by the vortex loops, and $\Psi$ fields are the neutral spin-1/2 excitations one can define near the nodes in the dSC. The coupling constant $e$ ("charge") is proportional to the dual order parameter that signals the appearance of infinitely large vortex loops, i.e. the destruction of the phase coherence in the dSC [1], [2]. We have assumed $N/2$ identical copies of each type of spinon field; the value of physical interest being N=2. Weak short-ranged interaction terms quartic in $\Psi$ are irrelevant by power counting and omitted in Eq. 1. The last term in Eq. 1 is the gauge-fixing term, and as customary we will work in the Landau gauge $\epsilon = 0$. Due to the $1/\sqrt{\lambda} \to \sqrt{\lambda}$ exchange in $\mathcal{L}_2$ it will prove necessary to keep both $\delta$ and $\lambda$ anisotropies in the action, once renormalization effects are included. In the limit $\delta \to 1$, $\lambda \to 1$ the model simply reduces to the much studied $N$-flavour $QED_3$.

In the calculation of the polarization tensor $\Pi_{\mu,\nu}$, for $\delta \neq 1$ and $\lambda \neq 1$, we use the bare spinon propagators and work only to leading order in $1/N$, as shown in Fig.1. In Appendix A we give explicit forms of $D_{\mu\nu}$ and $\Pi_{\mu\nu}$. The full gauge field propagator obeys the following symmetry properties:

$$D_{0,0}(p_0, p_1, p_2) = D_{0,0}(p_0, p_2, p_1)$$
$$D_{1,1}(p_0, p_1, p_2) = D_{2,2}(p_0, p_2, p_1)$$
$$D_{0,1}(p_0, p_1, p_2) = D_{0,2}(p_0, p_2, p_1)$$
$$D_{1,2}(p_0, p_1, p_2) = D_{1,2}(p_0, p_2, p_1). \quad (2)$$

In what follows, for the sake of clarity, we shall omit the calculational details and just outline our method. The interested reader is referred Appendix B where some of the more important details are given. We start this discussion by considering the form the full spinon propagator at the ith node should take

$$S_{i,R}(\vec{k}) = \left(\Sigma_i(\vec{k}) + A_{i,\mu}(\vec{k})\gamma_\mu k_\mu\right)^{-1}. \quad (3)$$

Here, the repeated index, as usual, implies a sum over indices. $\Sigma_i$ are the dynamically generated mass-gaps. We refer to $A_{i,\mu}$ as the wave-function renormalization in $k_\mu$ direction, for the spinon at the ith node. Using our form for the full gauge-field propagator and the above form for the spinon propagator we then proceed to construct the approximate Schwinger-Dyson equations for the full spinon propagators as shown diagrammatically in Fig.2.

From Eq. 2 and the form of the Schwinger-Dyson equations, it is easy to show that the following symmetry properties must also apply for the quantities we are interested in:

$$\Sigma_1(k_0, k_1, k_2) = \Sigma_2(k_0, k_2, k_1)$$
$$A_{1,i}(k_0, k_1, k_2) = A_{2,i}(k_0, k_2, k_1). \quad (4)$$

We therefore need consider only type 1 spinons and hereafter drop the spinon index. At this point it is also convenient for us to introduce the renormalized anisotropies $\lambda_R = A_2(0)/A_1(0)$ and $\delta_R = \sqrt{A_1(0)A_2(0)}/A_0(0)$.

From the Schwinger-Dyson equations we are able to generate four independent equations which we shall analyse. For the purposes of illustration, we will give here, only, the mass-gap equation. The other equations correspond to $A_0$, $A_1$ and $A_2$ from which we derive expressions for $\lambda_R$ and $\delta_R$; these are given in Appendix B. The mass equation takes the form

$$\Sigma(\vec{p}) = \frac{\alpha}{N} \int \frac{d^3 \vec{k}}{(2\pi)^3} \frac{\Sigma(\vec{k})}{U(\vec{k})} \mathcal{K}(\vec{q}), \quad (5)$$

where

$$U(\vec{k}) = k_0^2 + \lambda \delta^2 k_2^2 + \frac{\delta^2}{\lambda} k_1^2 + \Sigma_1(\vec{k})^2$$
$$\mathcal{K}(\vec{k}) = D_{0,0}(\vec{k}) + \frac{\delta^2}{\lambda} D_{1,1}(\vec{k}) + \delta^2 \lambda D_{2,2}(\vec{k}) \quad (6)$$

with $\vec{q} = \vec{k} - \vec{p}$, $\vec{k} = (k_0, k_1, k_2)$, and $\alpha = Ne^2$. These equations are difficult to use in their present form, due to the cumbersome form of $\mathcal{K}$. Therefore, we limit ourselves to the case of weak anisotropy, and treat $\delta' = \delta - 1$ and small $\lambda' = \lambda - 1$ as small parameters.

By expanding the mass-gap equation up to second order in both $\delta'$ and $\lambda'$ we may write $N_c$ in following form that depends on a rescaled self-energy $\Sigma_r(\vec{p}/\Sigma) = \Sigma(\vec{p})/\Sigma$, where $\Sigma = \Sigma(0)$,

$$N_c = \int \frac{d^3 k}{(2\pi)^3} \Sigma_r(\vec{k})[F^{0,0}(\vec{k}, \vec{0}) + \lambda' F^{1,0}(\vec{k}, \vec{0})$$
$$+\delta' F^{0,1}(\vec{k}, \vec{0}) + \delta' \lambda' F^{1,1}(\vec{k}, \vec{0})$$
$$+\delta'^2 F^{0,2}(\vec{k}, \vec{0}) + \lambda'^2 F^{0,2}(\vec{k}, \vec{0})] \quad (7)$$

where the forms for $F^{0,0}$-$F^{0,2}$ are given in Appendix B, but with $\Sigma(\vec{k})$ replaced with $\Sigma_r(\vec{k})$.

Next we further approximate $\Sigma_r(\vec{k})$ with its form in the isotropic limit, $\Sigma_r^0(k)$, consistent to leading order in the $1/N_c^0$ expansion which we employ. The $O(1/N_c^0)$ correction to $\Sigma_r$ is then given by

$$\Delta\Sigma_r(\vec{p}) = \int \frac{d^3 \vec{k}}{(2\pi)^3} \frac{\Sigma_r^0(k)}{N_c^0} [\lambda' F^{1,0}(\vec{k}, \vec{p}; \Sigma_r^0)$$
$$+\delta' F^{0,1}(\vec{k}, \vec{p}; \Sigma_r^0) + \delta' \lambda' F^{1,1}(\vec{k}, \vec{p}; \Sigma_r^0)$$
$$+\delta'^2 F^{0,2}(\vec{k}, \vec{p}; \Sigma_r^0) + \delta'^2 F^{2,0}(\vec{k}, \vec{p}; \Sigma_r^0)]. \quad (8)$$

By replacing $\Sigma_r^0(k)$ with $\Sigma_r^0(k) + \Delta\Sigma_r(\vec{k})$ in the equation for $N_c$ and expanding to first order in $\Delta\Sigma_r$ one is able to generate an $O(1/N_c^0)$ term in the expansion of $\Delta N_c$; which is finite. However, to be fully consistent in calculating the $O(1/N_c^0)$ correction in such a calculation, together with the correction $\Delta\Sigma$, corrections arising from the vertex and wave-function renormalization must be included. Indeed, it is well known that these change the value of $N_c^0$ even in the fully isotropic case [9]. In our calculation, all the terms that contribute to $O(1/N_c^0)$ are shown in Fig 3. Terms that contribute to $\Delta N_c$, generated from the vertex correction and wave-function renormalizations will contain logarithmic singularities. As we shall see, $\delta_R'$ and $\lambda_R'$ will also contain logarithmic singularities. So we assume we may absorb the singularities in $\Delta N_c$ into the renormalized quantities $\delta_R'$ and $\lambda_R'$, which would provide us with a well defined expansion.

We now calculate the leading order term in this expansion. It is easy to show that both the coefficients of $\lambda'$ and $\lambda'\delta'$ vanish under angular integration. We are then left with the problem of determining the $\Sigma_r^0$. One cannot, simply, set $\Sigma_r^0(k) = 1$, for one would obtain $N_c^0 = \infty$, contrary to the values obtained by direct solution of the mass-gap equation in the isotropic limit. $\Sigma_r^0(k)$ in fact must be a monotonicaly decreasing function of rescaled momentum. Based on the solution in [3] and the fact that that $\Sigma_r^0(0) = 1$, we propose a simple ansatz for $\Sigma_r^0(k)$ at $N_c$

$$\Sigma_r^0(k) = \frac{1}{1 + \frac{\sqrt{k}}{\Lambda}}. \tag{9}$$

By inserting this form into the mass-gap equation Eq. 5 in the isotropic limit one is able to find equations that fix $\Lambda$ and determine $N_c^0$. The points $p = 0$ and $p = \Lambda$ are only considered; at which we may set $\Sigma_r^0$, to 1 and 1/2, respectively. The solution to these equations is then

$\Lambda = 5.60$ and $N_c^0 = 3.19$. This is very close to $N_c^0 = 3.24$ [3], so the ansatz Eq. 9 will be sufficient for our purposes. We then use this to determine $\Delta N_c$ as a function of bare anisotropies:

$$\Delta N_c = 0.429\delta' + 0.566\delta'^2 + 0.406\lambda'^2. \tag{10}$$

Furthermore, we are able to show that both the $\delta'$ and the $\lambda'^2$ coefficients are positive definite, for all possible forms of $\Sigma_r^0(k)$. Although we are unable to prove this generally for the $\delta'^2$ term, we have been able to show that it remains positive over all chosen values of the parameter $\Lambda$; not just for $\Lambda = 5.60$.

As argued above, we then take $\Delta N_c$ to be given by the above expression, but with the bare anisotropies replaced with the renormalized values. Since the latter differ from the bare ones by terms of $O(1/N)$, this is certainly correct to the zeroth order in the expansion in $1/N$. We are assuming, however, that by doing this we are effectively including the leading logarithmic divergences from the vertex corrections in Fig. 3., which are formally of the next order in $1/N$. One justification for this assumption will be that by doing this we will find that $N_c$ becomes independent of the parameter $\delta$; as it should be, since we could have rescaled the theory so to place $\delta$ entirely into the gauge-field propagator and the gauge fixing term. Since the former is irrelevant compared to the fermion polarization, gauge invariance ensures any dependence on $\delta$ in Eq. 1 does not effect physical quantities. The other justification is that one could expect a quantity like $N_c$ to be universal; i.e. independent of all weak irrelevant perturbations. In the example of the chiral instability in presence of weak quartic interactions, which are irrelevant $N_c$, indeed, does not change [10]. It is also easy to check that in the isotropic $QED_3$ including higher derivatives in the fermionic or gauge fields can not affect $N_c^0$, to the leading order in $1/N$. If, indeed, $N_c$ is universal we can fully replace the bare anisotropies with renormalized ones and show that $N_c$ will be independent of anisotropy, since we will ultimately find anisotropy to be an irrelevant perturbation.

We therefore next determine $\lambda_R = 1 + \lambda_R'$ and $\delta_R = 1 + \delta_R'$ from our approximate Schwinger-Dyson equations. For the equations for $A$, $\delta_R'$ and $\lambda_R'$ we find forms similar to that of Eq. 8, when we expand in powers of $\delta'$ and $\lambda'$. In the limit $\Sigma/\alpha \to 0$, certain terms develop singularities, which dominate the behaviour of $A_0$, $\delta_R$ and $\lambda_R$. Expressed in terms of these singularities we find

$$A_0 = 1 + \frac{\hat{A}\delta'}{N} - \frac{8}{3\pi^2 N}\left(1 + \left(\frac{4}{7} - \frac{4}{5}\right)\delta'\right)\ln\left(\frac{\alpha}{\Sigma}\right)$$

$$\delta_R' = 1 + \left(\frac{\hat{B}}{N} - \frac{8}{3\pi^2 N}\left(\frac{4}{5} - \frac{4}{7}\right)\ln\left(\frac{\alpha}{\Sigma}\right)\right)\delta'$$

$$\lambda_R' = 1 + \left(\frac{\hat{C}}{N} - \frac{32}{5\pi^2 N}\ln\left(\frac{\alpha}{\Sigma}\right)\right)\lambda'. \tag{11}$$

$\hat{A}$, $\hat{B}$, and $\hat{C}$, are finite numerical constants. From these we may read off the flow of anisotropies to order $O(1/N)$:

$$\begin{aligned}\frac{d\delta_R'}{ds} &= -\frac{64}{105\pi^2 N}\delta_R', \\ \frac{d\lambda_R'}{ds} &= -\frac{32}{5\pi^2 N}\lambda_R',\end{aligned} \tag{12}$$

where $s = \ln(\alpha/\Sigma)$. Weak $\delta_R'$ and $\lambda_R'$ are thus both irrelevant as $\Sigma \to 0$ and flow to zero. This agrees with the conclusion of [11]. From this and the Eq. 9 we deduce that to leading order in $1/N_c^0$, $\Delta N_c = 0$.

One should emphasise, for the above conclusion to hold to all orders in our expansion in $N_c$ it appears essential that $N_c$ is written in terms of renormalized anisotropies only, which would render it a universal number, dependent only on the infrared behaviour in the theory. Although one would expect this to be true in all cases, based on the examples given previously; it is by no means obvious that our expectation is true in the case of anisotropy, or for that matter, other marginal perturbations.

To summarize, we find that $\Delta N_c = 0$ to leading order in $1/N_c^0$. This implies that weak velocity anisotropy does not effect $N_c$ and that one is free to set $v_F = v_\Delta = c = 1$ in the determination of all critical quantities. However, issues such as the universality of $N_c$ and the regime of strong anisotropy, need to be considered before one is certain that this is, indeed, the case beyond the limits of our calculation. We hope to address such problems in future work.

This work is supported by NSERC of Canada and the Research Corporation.

# APPENDIX A: FULL GAUGE FIELD PROPAGATOR

To calculate the full gauge field propagator, we first need to calculate the polaziation function given in Fig. 1. We first note that from the Feynman rules

$$\Pi_{\mu,\nu}(\vec{p}) = \sum_i^2 \int \frac{d^3k}{(2\pi)^3} \tilde{\gamma}_{\mu,i} S_{i,0}(\vec{k}) \tilde{\gamma}_{\nu,i} S_0(\vec{k}+\vec{p}) \quad (A1)$$

where $\tilde{\gamma}_{0,1} = \gamma_0$, $\tilde{\gamma}_{1,1} = \delta/\sqrt{\lambda}\gamma_1$, and $\tilde{\gamma}_{2,1} = \delta\sqrt{\lambda}\gamma_2$; and also $\tilde{\gamma}_{\mu,2}(\lambda) = \tilde{\gamma}_{\mu,2}(1/\lambda)$. $S_{i,0}$ is the bare spinon propagator for the ith node, and takes the form

$$S_{i,0} = \frac{1}{\tilde{\gamma}_{\mu,i} p_\mu}. \quad (A2)$$

Here, as usual, a repeated index implies a sum over that index. It is easy to deduce from the calculation of $\Pi_{\mu,\nu}$ in the isotropic limit what explicit form the elements of $\Pi_{\mu,\nu}$ should take

$$\Pi_{0,0} \equiv \bar{\Pi}_1(p_0,p_1,p_2) =$$
$$\frac{1}{8}\left[\frac{1}{P_1}\left(\frac{p_1^2}{\lambda}+\lambda p_2^2\right)+\frac{1}{P_2}\left(\lambda p_1^2+\frac{p_2^2}{\lambda}\right)\right]$$
$$\Pi_{1,1}(\vec{p}) \equiv \Pi_{2,2}(\vec{p}_E) \equiv \bar{\Pi}_2(\vec{p}) =$$
$$\frac{1}{8}\left[\frac{1}{\lambda P_1}\left(p_0^2+\lambda p_2^2\right)+\frac{\lambda}{P_2}\left(p_0^2+\frac{p_2^2}{\lambda}\right)\right]$$
$$\Pi_{1,0}(\vec{p}) \equiv \Pi_{2,0}(\vec{p}_E) \equiv p_0 p_1 \bar{\Pi}_3(\vec{p}) =$$
$$-\frac{p_0 p_1}{8}\left[\frac{1}{\lambda P_1}+\frac{\lambda}{P_2}\right]$$
$$\Pi_{1,2}(\vec{p}) \equiv -p_1 p_2 \bar{\Pi}_4(\vec{p}) =$$
$$\frac{p_1 p_2 \delta^2}{8}\left[\frac{1}{P_1}+\frac{1}{P_2}\right]. \quad (A3)$$

Here, for convenience, we employ the notation: $\vec{p} = (p_0,p_1,p_2)$, $\vec{p}_E = (p_0,p_2,p_1)$, $P_1 = (p_0^2+\frac{\delta^2}{\lambda}p_1^2+\delta^2\lambda p_2^2)^{1/2}$ and $P_2(\lambda) = P_1(1/\lambda)$. One may immeadiately write down the expression for the full inverse gauge field propagator

$$D^{-1}_{\mu,\nu} = k^2\left(\delta_{\mu\nu}-(1-\epsilon)\frac{k_\mu k_\nu}{k^2}\right)+\frac{\alpha}{2}\Pi_{\mu\nu} \quad (A4)$$

Straightforwardly although awkwardly, from Eq. A.4 one is able to invert the inverse gauge field propagator, and so obtain expressions for the elements of the gauge-field propagator for arbitrarily covariant gauge.

$$D_{0,0}(\vec{p}) \equiv \bar{D}_1(\vec{p}) \quad D_{1,1}(\vec{p}) = D_{2,2}(\vec{p}_E) \equiv \bar{D}_2(\vec{p})$$
$$D_{0,1}(\vec{p}) = D_{0,2}(\vec{p}_E) \equiv \bar{D}_3(\vec{p}) \quad D_{1,2}(\vec{p}) \equiv \bar{D}_4(\vec{p})$$
$$\bar{D}_i(\vec{p}) = \mathcal{P}_\epsilon(\vec{p})(D_{i,L}(\vec{p})+\epsilon D_{i,\epsilon}(\vec{p}))$$
$$(A5)$$

where

$$D_{1,L}(\vec{p}) = \left[p_1^2\left(p^2+\frac{\alpha}{2}\bar{\Pi}_2(\vec{p}_E)\right)+\right.$$
$$\left. p_2^2\left(p^2+\frac{\alpha}{2}\bar{\Pi}_2(\vec{p})\right)-\alpha p_1^2 p_2^2 \bar{\Pi}_4(\vec{p})\right]$$
$$D_{1,\epsilon}(\vec{p}) = \left(p_0^2+p_1^2+\frac{\alpha}{2}\bar{\Pi}_2(\vec{p}_E)\right)$$
$$\times\left(p_0^2+p_2^2+\frac{\alpha}{2}\bar{\Pi}_2(\vec{p})\right)-p_1^2 p_2^2\left(\frac{\alpha}{2}\bar{\Pi}_4(\vec{p})-1\right)^2$$
$$D_{2,L}(\vec{p}) = \left[p_0^2\left(p^2+\frac{\alpha}{2}\bar{\Pi}_2(\vec{p}_E)\right)+\right.$$
$$\left. p_2^2\left(p^2+\frac{\alpha}{2}\bar{\Pi}_1(\vec{p})\right)-\alpha p_0^2 p_2^2 \bar{\Pi}_3(\vec{p}_E)\right]$$
$$D_{2,\epsilon}(\vec{p}) = \left(p_0^2+p_1^2+\frac{\alpha}{2}\bar{\Pi}_2(\vec{p}_E)\right)$$
$$\times\left(p_1^2+p_2^2+\frac{\alpha}{2}\bar{\Pi}_1(\vec{p})\right)-p_0^2 p_2^2\left(\frac{\alpha}{2}\bar{\Pi}_3(\vec{p}_E)-1\right)^2$$
$$D_{3,L}(\vec{p}) = \left[p_2^2\frac{\alpha}{2}\left(\bar{\Pi}_3(\vec{p}_E)+\bar{\Pi}_4(\vec{p})-\bar{\Pi}_3(\vec{p}_E)\right)\right.$$
$$\left. -\left(p^2+\frac{\alpha}{2}\Pi_2(\vec{p}_E)\right)\right]p_0 p_1$$
$$D_{3,\epsilon}(\vec{p}) = p_0 p_1\left[p_2^2\left(\frac{\alpha}{2}\bar{\Pi}_3(\vec{p}_E)-1\right)\left(\frac{\alpha}{2}\bar{\Pi}_4(\vec{p})-1\right)\right.$$
$$\left. -\left(\frac{\alpha}{2}\bar{\Pi}_3(\vec{p})-1\right)\left(p_0^2+p_1^2+\frac{\alpha}{2}\bar{\Pi}_2(\vec{p}_E)\right)\right]$$
$$D_{4,L}(\vec{p}) = \left[p_2^2\frac{\alpha}{2}\left(\bar{\Pi}_3(\vec{p}_E)+\bar{\Pi}_3(\vec{p})-\bar{\Pi}_4(\vec{p})\right)\right.$$
$$\left. -\left(p^2+\frac{\alpha}{2}\bar{\Pi}_1(\vec{p})\right)\right]p_1 p_2$$
$$D_{4,\epsilon}(\vec{p}) = p_0 p_1\left[p_0^2\left(\frac{\alpha}{2}\bar{\Pi}_3(\vec{p})-1\right)\left(\frac{\alpha}{2}\bar{\Pi}_3(\vec{p}_E)-1\right)\right.$$
$$\left. -\left(\frac{\alpha}{2}\bar{\Pi}_4(\vec{p})-1\right)\left(p_1^2+p_2^2+\frac{\alpha}{2}\bar{\Pi}_1(\vec{p})\right)\right], (A6)$$

and

$$\mathcal{P}_\epsilon(\vec{p}) = \frac{p_0^2+\epsilon(p_1^2+p_2^2+\bar{\Pi}_1(\vec{p}))}{\mathcal{D}_L(\vec{p})+\epsilon\mathcal{D}_{\epsilon,1}(\vec{p})+\epsilon^2\mathcal{D}_{\epsilon,2}(\vec{p})} \quad (A7)$$

where

$$\mathcal{D}_L(\vec{p}) = A_L(\vec{p})A_L(\vec{p}_E)-B_L(\vec{p})^2$$
$$\mathcal{D}_{\epsilon,1}(\vec{p}) = A_L(\vec{p})A_\epsilon(\vec{p}_E)$$
$$+A_L(\vec{p}_E)A_\epsilon(\vec{p})-2B_L(\vec{p})B_\epsilon(\vec{p})$$
$$\mathcal{D}_{\epsilon,2}(\vec{p}) = A_\epsilon(\vec{p})B_\epsilon(\vec{p})-B_\epsilon(\vec{p})^2 \quad (A8)$$

and

$$A_L(\vec{p}) = \left[p_0^2\left(p^2+\frac{\alpha}{2}\bar{\Pi}_2(\vec{p})\right)\right.$$
$$\left. +p_1^2\left(p^2+\frac{\alpha}{2}\bar{\Pi}_1(\vec{p})\right)-\alpha p_0^2 p_1^2 \bar{\Pi}_3(\vec{p})\right]$$
$$B_L(\vec{p}) = p_1 p_2\left[\left(p^2+\frac{\alpha}{2}\bar{\Pi}_1(\vec{p})\right)\right.$$
$$\left. +p_0^2\frac{\alpha}{2}\left(\bar{\Pi}_4(\vec{p})-\bar{\Pi}_3(\vec{p})-\bar{\Pi}_3(\vec{p}_E)\right)\right]$$
$$A_\epsilon(\vec{k}) = \left[\left(p_0^2+p_2^2+\frac{\alpha}{2}\bar{\Pi}_2(\vec{k})\right)\right.$$
$$\left.\left(p_1^2+p_2^2+\frac{\alpha}{2}\bar{\Pi}_1(\vec{k})\right)-p_0^2 p_1^2\left(\frac{\alpha}{2}\bar{\Pi}_3(p_0,p_1,p_2)-1\right)^2\right]$$
$$B_\epsilon(\vec{p}) = p_1 p_2\left[\left(\frac{\alpha}{2}\bar{\Pi}_4(\vec{p})-1\right)\left(p_1^2+p_2^2+\frac{\alpha}{2}\bar{\Pi}_1(\vec{p})\right)\right.$$
$$\left. -p_0^2\left(\frac{\alpha}{2}\bar{\Pi}_3(\vec{p})-1\right)\left(\frac{\alpha}{2}\bar{\Pi}_3(\vec{p}_E)-1\right)\right].$$
$$(A9)$$

# APPENDIX B: DETAILS IN DERIVING THE FORMULA FOR $N_c$, $\delta_R$ AND $\lambda_R$ FROM SCHWINGER-DYSON EQUATIONS

In what follows we shall limit ourselves to the Landau gauge. From Fig. 2 we may easily write down the expressions for the S-D equations

$$S_{i,R}(\vec{p})^{-1} = S_{i,0}(\vec{p})^{-1} - e^2 \int \frac{d^3k}{(2\pi)^3} \tilde{\gamma}_{\mu,i} S_{i,D}(\vec{k}) \tilde{\gamma}_{\nu,i} D_{\mu,\nu}(\vec{q}) \tag{B1}$$

Where $S_{i,R}$ and $S_{i,D}$ are

$$S_{i,R}(\vec{k}) = \left(\Sigma_i(\vec{k}) + A_{i,\mu}(\vec{k})\gamma_\mu k_\mu\right)^{-1}$$
$$S_{i,D}(\vec{k}) = \left(\Sigma_i(\vec{k}) + \tilde{\gamma}_{\mu,i} p_\mu\right)^{-1}. \tag{B2}$$

By performing the trace of Eq. B.1 we arrive at the mass-gap equations

$$\Sigma_i(\vec{p}) = \frac{\alpha}{N} \int \frac{d^3\vec{k}}{(2\pi)^3} \frac{\Sigma_i(\vec{k})}{U_i(\vec{k})} \mathcal{K}_i(\vec{q}), \tag{B3}$$

where

$$U_1(\vec{k}) = k_0^2 + \lambda \delta^2 k_2^2 + \frac{\delta^2}{\lambda} k_1^2 + \Sigma_1(\vec{k})^2$$
$$U_2(\vec{k}) = k_0^2 + \frac{\delta^2}{\lambda} k_2^2 + \delta^2 \lambda k_1^2 + \Sigma_2(\vec{k})^2$$
$$\mathcal{K}_1(\vec{k}) = D_{0,0}(\vec{k}) + \frac{\delta^2}{\lambda} D_{1,1}(\vec{k}) + \delta^2 \lambda D_{2,2}(\vec{k})$$
$$\mathcal{K}_2(\vec{k}) = D_{0,0}(\vec{k}) + \delta^2 \lambda D_{1,1}(\vec{k}) + \frac{\delta^2}{\lambda} D_{2,2}(\vec{k}). \tag{B4}$$

By using the symmetry properties given by Eq. 2, in the text, it is very easy to show that, indeed, $\Sigma_1(k_0, k_1, k_2) = \Sigma(k_0, k_2, k_1)$. By multiplying Eq. B.1 from the left with $\gamma_0$ and then taking the trace we arrive at the equation for $A_{1,0}$

$$A_{1,0}(\vec{p}) = 1 - \frac{\alpha}{Np_0} \int \frac{d^3\vec{k}}{(2\pi)^3} \frac{2\mathcal{T}_0(\vec{k},\vec{q}) - k_0 \mathcal{K}(\vec{q})}{U_1(\vec{k})}, \tag{B5}$$

where

$$\mathcal{T}_i(\vec{k},\vec{q}) = k_0 D_{i,0}(\vec{q}) + \frac{\delta^2 k_1}{\lambda} D_{i,1}(\vec{q}) + \delta^2 \lambda k_2 D_{i,2}(\vec{q}). \tag{B6}$$

Similarly, by multiplying Eq. B.1 by either $\gamma_1$ or $\gamma_2$ one obtains equations for $A_{1,1}$ and $A_{1,2}$, respectively

$$A_{1,1}(\vec{p}) = \frac{\delta}{\sqrt{\lambda}} \left(1 - \frac{\alpha}{Np_1} \int \frac{d^3k}{(2\pi)^3} \frac{2\mathcal{T}_1(\vec{k},\vec{q}) - k_1 \mathcal{K}(\vec{q})}{U(\vec{k})}\right),$$
$$A_{1,2}(\vec{p}) = \delta\sqrt{\lambda} \left(1 - \frac{\alpha}{Np_2} \int \frac{d^3k}{(2\pi)^3} \frac{2\mathcal{T}_2(\vec{k},\vec{q}) - k_2 \mathcal{K}(\vec{q})}{U(\vec{k})}\right) \tag{B7}$$

Again, it is easy to show from the symmetry properties of $D_{\mu,\nu}$ that $A_{1,i}(k_0, k_1, k_2) = A_{2,i}(k_0, k_2, k_1)$. For the moment we will concentrate on the mass-gap equation and calculate $N_c$, before looking at wave-function renormalization.

Now, as discussed in the text, we perform an expansion in both $\delta' = \delta - 1$ and $\lambda' = \lambda - 1$ which gives us up to second order

$$\Sigma(\vec{p}) = \frac{1}{N} \int \frac{d^3k}{(2\pi)^3} \Sigma(\vec{k}) [F^{0,0}(\vec{k},\vec{p}) + \lambda' F^{1,0}(\vec{k},\vec{p})$$
$$+ \delta' F^{0,1}(\vec{k},\vec{p}) + \delta'\lambda' F^{1,1}(\vec{k},\vec{p})$$
$$+ \delta'^2 F^{0,2}(\vec{k},\vec{p}) + \lambda'^2 F^{0,2}(\vec{k},\vec{p})] \tag{B8}$$

where

$$F^{0,0}(\vec{k},\vec{p}) = \frac{16}{q} \frac{1}{I(\vec{k})},$$
$$F^{1,0} = \frac{16K_-}{I^2 q} + \frac{8Q_-}{q^3 I},$$
$$F^{0,1} = \frac{16Q_0}{Iq^3} - \frac{32K_+}{I^2 q},$$
$$F^{1,1} = \left(\frac{2K_-}{I^2} - \frac{4K_+ K_-}{I^3}\right) \frac{16}{q}$$
$$+ \frac{8K_- Q_0 - 16K_+ Q_-}{I^2 q^3} + \frac{8q^2 Q_- - 8q_0^2 Q_-}{q^5 I},$$
$$F^{2,0} = \left(\frac{K_-^2}{I^3} - \frac{k_1^2}{I^2}\right) \frac{16}{q} + \frac{K_- Q_-}{I^2 q^3} + \frac{(8q_1^2 q^2 - 2Q_-^2)}{q^5 I},$$
$$F^{0,2} = \left(\frac{4K_+^2}{I^3} - \frac{K_+}{I^2}\right) \frac{16}{q} - \frac{32K_+ Q_0}{Iq^3} + \frac{8q^2 Q_0 - 8Q_+^2}{q^5 I} \tag{B9}$$

where $K_+ = k_1^2 + k_2^2$, $K_- = k_1^2 - k_2^2$, $Q_+ = q_1^2 + q_2^2$, $Q_- = q_1^2 - q_2^2$, $Q_0 = q_0^2 + q^2$, $I(\vec{k}) = k^2 + \Sigma(\vec{k})^2$ and $\vec{q} = \vec{k} - \vec{p}$. By rescaling the self-energy $\Sigma_r(\vec{p'}) = \Sigma(\vec{p})/\Sigma$, loop momenta $k' = k/\Sigma$; and setting external momenta $p = 0$ we arrive at Eq. 7 given in the text. By performing the expansion in $\delta'$ and $\lambda'$ to first order we find at an expression for $A_0$

$$A_0(\vec{p}) = 1 - \frac{1}{Np_0} \int \frac{d^3k}{(2\pi)^3} (\tilde{F}_0^{0,0}(\vec{k},\vec{p}) - k_0 F^{0,0}(\vec{k},\vec{p}))$$
$$+ \lambda'(\tilde{F}_0^{1,0}(\vec{k},\vec{p}) - k_0 F^{1,0}(\vec{k},\vec{p}))$$
$$+ \delta'(\tilde{F}_0^{0,1}(\vec{k},\vec{p}) - k_0 F^{0,1}(\vec{k},\vec{p})) \tag{B10}$$

where

$$\tilde{F}_0^{0,0}(\vec{k},\vec{p}) = \frac{(k_0 Q_+ - q_0(q_1 k_1 + q_2 k_2))\,16}{q^3 I(\vec{k})},$$
$$\tilde{F}_0^{1,0} = \frac{16K_-(k_0 Q_+ - q_0(q_1 k_1 + q_2 k_2))}{I^2 q^3}$$
$$- \frac{1}{I} \frac{16q_0(k_2 q_2 - k_1 q_1)}{q^3},$$
$$\tilde{F}_0^{0,1} = \left(\frac{k_0 Q_+^2 + (q_1 k_1 - q_2 k_2) Q_- q_0}{q^5}\right.$$

$$-\frac{2(k_1q_1+k_2q_2)q_0}{q^3}\Bigg)\frac{16}{I}$$
$$-\frac{32K_+\left(k_0Q_+-q_0(q_1k_1+q_2k_2)\right)}{I^2q^3},\tag{B11}$$

For $A_1$ we find
$$A_1(\vec{p})=\left(1-\frac{1}{Np_1}\int\frac{d^3k}{(2\pi)^3}(\tilde{F}_1^{0,0}(\vec{k},\vec{p})-k_1F^{0,0}(\vec{k},\vec{p}))\right.$$
$$+\lambda'(\tilde{F}_1^{1,0}(\vec{k},\vec{p})-k_1F^{1,0}(\vec{k},\vec{p}))$$
$$\left.+\delta'(\tilde{F}_1^{0,1}(\vec{k},\vec{p})-k_1F^{0,1}(\vec{k},\vec{p}))\right)\frac{\delta}{\sqrt{\lambda}}\tag{B12}$$

where
$$\tilde{F}_1^{0,0}(\vec{k},\vec{p})=\frac{\left(k_1Q_+^{02}-q_1(q_0k_0+q_2k_2)\right)16}{q^3I(\vec{k})},$$
$$\tilde{F}_1^{1,0}=\frac{16K_-\left(k_1Q_+^{02}-q_1(q_0k_0+q_2k_2)\right)}{I^2q^3}$$
$$-\frac{16\left(k_1Q_+^{02}+k_2q_1q_2\right)}{Iq^3},$$
$$\tilde{F}_1^{0,1}=\left[\frac{k_2q_1q_2Q_+}{q^5}-\frac{k_0Q_-q_0q_1}{q^5}-\frac{k_1\left(Q_+Q_+^{02}-2q_0^2q_1^2\right)}{q^5}\right.$$
$$\left.+\frac{2\left(k_1Q_+^{02}-k_2q_1q_2\right)}{q^3}\right]\frac{16}{I}$$
$$-\frac{32K_+\left(k_1Q_+^{02}-q_1(q_0k_0+q_2k_2)\right)}{I^2q^3},\tag{B13}$$

where $Q_+^{02}=q_0^2+q_2^2$. For $A_2$ we find
$$A_1(\vec{p})=\left(1-\frac{1}{Np_2}\int\frac{d^3k}{(2\pi)^3}(\tilde{F}_2^{0,0}(\vec{k},\vec{p})-k_2F^{0,0}(\vec{k},\vec{p}))\right.$$
$$+\lambda'(\tilde{F}_2^{1,0}(\vec{k},\vec{p})-k_2F^{1,0}(\vec{k},\vec{p}))$$
$$\left.+\delta'(\tilde{F}_2^{0,1}(\vec{k},\vec{p})-k_2F^{0,1}(\vec{k},\vec{p}))\right)\delta\sqrt{\lambda}\tag{B14}$$

where
$$\tilde{F}_2^{0,0}(\vec{k},\vec{p})=\frac{\left(k_2Q_+^{01}-q_2(q_0k_0+q_1k_1)\right)16}{q^3I(\vec{k})},$$
$$\tilde{F}_2^{1,0}=-\frac{16K_-\left(k_2Q_+^{01}-q_2(q_0k_0+q_1k_1)\right)}{I^2q^3}$$
$$+\frac{16\left(k_2Q_+^{01}+k_1q_2q_1\right)}{Iq^3},$$
$$\tilde{F}_1^{0,1}=\left[\frac{k_1q_1q_2Q_+}{q^5}-\frac{k_0Q_-q_0q_2}{q^5}-\frac{k_2\left(Q_+Q_+^{01}-2q_0^2q_2^2\right)}{q^5}\right.$$
$$\left.+\frac{2\left(k_2Q_+^{01}-k_1q_1q_2\right)}{q^3}\right]\frac{16}{I}$$
$$-\frac{32K_+\left(k_2Q_+^{01}-q_2(q_0k_0+q_1k_1)\right)}{I^2q^3},\tag{B15}$$

where $Q_+^{01}=q_0^2+q_1^2$.

Now, we are interested in calculating the singular contributions to $\lambda_R$ and $\delta_R$ in the limit $\Sigma\to 0$. To do this we require the singular contributions to $A_0(\vec{0})$, $A_1(\vec{0})$ and $A_2(\vec{0})$. In finding the logarithmically singular contributions one may neglect the momentum dependence of $\Sigma(\vec{p})$ and simply replace it with $\Sigma(\vec{0})$. Then one is able to perform the angular integrations. We find the following

$$A_0\approx 1-\frac{8}{3\pi^2N}\left[1+\left(\frac{4}{7}-\frac{4}{5}\right)\delta'\right]\ln\left(\frac{\alpha}{\Sigma}\right)$$
$$A_1\approx\left(1-\frac{8}{3\pi^2N}\left[1-\frac{6\lambda'}{5}\right]\ln\left(\frac{\alpha}{\Sigma}\right)\right)\frac{\delta}{\sqrt{\lambda}}$$
$$A_2\approx\left(1-\frac{8}{3\pi^2N}\left[1+\frac{6\lambda'}{5}\right]\ln\left(\frac{\alpha}{\Sigma}\right)\right)\delta\sqrt{\lambda}.\tag{B16}$$

Where, we have neglected non-singular $O(1/N)$ terms. Using the definitions for $\delta_R$ and $\lambda_R$ given in the text, we thereby obtain Eq. 11.

---

FIG. 1: The polarisation tensor $\Pi_{\mu,\nu}$ to the leading order in $1/N$. The thin straight lines with the numbers one and two represent bare spinon 1 and 2 propagators.

FIG. 2: The Schwinger-Dyson equations. The full spinon propagators are represented by thick lines. Spinon propagators that contain unrenormalized anisotropy ($\delta,\lambda$), but the full mass gap, are represented by double lines. The full gauge-field propagator is represented by a wavy line.

FIG. 3: Graphs that contribute to the $O(1/N_c^0)$ correction to $\Delta N_c$. Closed spinon loops imply a sum over both nodes.